\documentclass{cernyrep} 
\usepackage{texnames}
\usepackage[T1]{fontenc}
\usepackage{comment}

\usepackage{tikz}
\usetikzlibrary{arrows,shapes}
\usetikzlibrary{trees}
\usetikzlibrary{matrix,arrows} 				
\usetikzlibrary{positioning}				
\usetikzlibrary{calc,through}				
\usetikzlibrary{decorations.pathreplacing}  
\usepackage{pgffor}							

\usetikzlibrary{decorations.pathmorphing}	
\usetikzlibrary{decorations.markings}
\usetikzlibrary{snakes}
\tikzset{
    vector/.style={decorate, decoration={snake}, draw},
	provector/.style={decorate, decoration={snake,amplitude=2.5pt}, draw},
	antivector/.style={decorate, decoration={snake,amplitude=-2.5pt}, draw},
    fermion/.style={draw=black, postaction={decorate},
        decoration={markings,mark=at position .55 with {\arrow[draw=black]{>}}}},
    fermionbar/.style={draw=black, postaction={decorate},
        decoration={markings,mark=at position .55 with {\arrow[draw=black]{<}}}},
    fermionnoarrow/.style={draw=black},
    gluon/.style={decorate, draw=black,
        decoration={coil,amplitude=4pt, segment length=5pt}},
    scalar/.style={dashed,draw=black, postaction={decorate},
        decoration={markings,mark=at position .55 with {\arrow[draw=black]{>}}}},
    scalarbar/.style={dashed,draw=black, postaction={decorate},
        decoration={markings,mark=at position .55 with {\arrow[draw=black]{<}}}},
    scalarnoarrow/.style={dashed,draw=black},
    electron/.style={draw=black, postaction={decorate},
        decoration={markings,mark=at position .55 with {\arrow[draw=black]{>}}}},
	bigvector/.style={decorate, decoration={snake,amplitude=4pt}, draw},
}

\tikzstyle{block} = [draw, rectangle, 
    minimum height=3em, minimum width=6em]

\begin{document}
\title{Less is More when Gluinos Mediate}

\author{Benjamin Nachman}

\institute{SLAC, Stanford University}

\maketitle 

\begin{abstract}
Compressed mass spectra are generally more difficult to identify than spectra with large splittings.   In particular, gluino pair production with four high energy top or bottom quarks leaves a striking signature in a detector.  However, if any of the mass splittings are compressed, the power of traditional techniques may deteriorate.   Searches for direct stop/sbottom pair production can fill in the gaps.   As a demonstration, we show that for $\tilde{g}\rightarrow t\tilde{t}_1$ and $m_{\tilde{t}_1}\sim m_{\tilde{\chi}_1^0}$, limits on the stop mass at 8 TeV can be extended by least 300 GeV for a 1.1 TeV gluino using a $pp\rightarrow \tilde{t}_1\tilde{t}_1$ search.  At 13 TeV, the effective cross section for the gluino mediated process is twice the direct stop pair production cross section, suggesting that direct stop searches could be sensitive to discover new physics earlier than expected.
\end{abstract}
 
\section{Introduction} 


With the recent discovery of the Higgs boson with a mass of 125 GeV by the ATLAS~\cite{Aad:2012tfa} and CMS~\cite{Chatrchyan:2012ufa} collaborations at the LHC, the issue of naturalness is brought into focus.    If the solution to the hierarchy problem is that the Standard Model is a subset of a supersymmetric (SUSY) theory, then naturalness suggests that the mass of the supersymmetric counterpart to the top quark, the stop ($\tilde{t}$), is near the electroweak scale $\sim \mathcal{O}(100)$ GeV.   For various reasons such as gauge unification or the 2-loop radiative corrections to the Higgs boson mass, one may also expect a light supersymmetric counterpart of the gluon, the gluino ($\tilde{g}$), with mass $\sim\mathcal{O}(1)$ TeV.   The spectrum can remain natural even if all other colored super partners have masses well into the TeV range or beyond.  In $R$-parity conserving SUSY theories, gluinos would be pair produced; the gluino pair production cross section is much larger than the cross section for direct pair produced stops.   For example, at 8 TeV, the cross section for pair produced stops with mass 800 GeV is about 0.002 pb whereas the cross section for stops produced from the decay of 1 TeV pair produced gluinos is about 0.02 pb~\cite{Kramer:2012bx}.   Both ATLAS~\cite{Aad:2013wta,Aad:2014lra,Aad:2014pda,Aad:2014wea,Aad:2015mia} and CMS~\cite{Chatrchyan:2014lfa,Chatrchyan:2013iqa,Chatrchyan:2013fea,CMS-PAS-SUS-14-011,CMS-PAS-SUS-13-016,CMS-PAS-SUS-13-008} have searched extensively for the scenario of {\it gluino mediated stop} production at the LHC and have excluded natural spectra with large mass splittings that have gluinos with mass below about 1.5 TeV.

\begin{figure}[h!]

\begin{center}
\begin{tikzpicture}[line width=1.5 pt, scale=1.3]
	
	\draw (-1,1+0.1)--(0,0+0.1);
	\draw (-1,-1+0.1)--(0,0+0.1);
	
	\draw (-1,1)--(0,0);
	\draw (-1,-1)--(0,0);
	
	\draw (-1,1-0.1)--(0,0-0.1);
	\draw (-1,-1-0.1)--(0,0-0.1);		
	
	\draw[gluon,color=red] (0,0)--(1,1);
	\draw[gluon,color=red] (0,0)--(1,-1);
	\draw[color=red] (0,0)--(1,1);
	\draw[color=red] (0,0)--(1,-1);

	\draw[dashed,color=red] (1,1)--(2,1.5);
	\draw[dashed,color=red] (1,-1)--(2,-1.5);

	\draw[] (1,1)--(2,0.5);
	\draw[] (1,-1)--(2,-0.5);

	\draw[color=red] (2,1.5)--(3,2);
	\draw[color=red] (2,-1.5)--(3,-2);
	\draw[vector,color=red] (2,1.5)--(3,2);
	\draw[vector,color=red] (2,-1.5)--(3,-2);	
	\draw[] (2,1.5)--(3,1);
	\draw[] (2,-1.5)--(3,-1);
	
	\draw[fill=black!30!white] (0,0) circle (.25);	
	
	\node at (1.5, 1.6) {\color{red} $\tilde{t}$};
	\node at (1.5, -1.6) {\color{red} $\tilde{t}$};
	\node at (2.5, 2.1) {\color{red} $\tilde{\chi}^0$};
	\node at (2.5, -2.1) {\color{red} $\tilde{\chi}^0$};	
	\node at (0.4, 0.8) {\color{red} $\tilde{g}$};
	\node at (0.4, -0.8) {\color{red} $\tilde{g}$};	
	\node at (2.2, 0.4) {$\bar{t}$};
	\node at (2.2, -0.4) {$\bar{t}$};
	\node at (3.4, 0.9) {\it soft};
	\node at (3.4, -0.9) {\it soft};
	
	\draw[implies-implies,double equal sign distance] (4.5,0) -- (5,0);
	
	 \begin{scope}[shift={(7,0)}]

	\draw (-1,1+0.1)--(0,0+0.1);
	\draw (-1,-1+0.1)--(0,0+0.1);
	
	\draw (-1,1)--(0,0);
	\draw (-1,-1)--(0,0);
	
	\draw (-1,1-0.1)--(0,0-0.1);
	\draw (-1,-1-0.1)--(0,0-0.1);		
	
	\draw[dashed,color=red] (0,0)--(1,1);
	\draw[dashed,color=red] (0,0)--(1,-1);

	\draw[vector,color=red] (1,1)--(2,1.5);
	\draw[vector,color=red] (1,-1)--(2,-1.5);
	\draw[color=red] (1,1)--(2,1.5);
	\draw[color=red] (1,-1)--(2,-1.5);

	\draw[] (1,1)--(2,0.5);
	\draw[] (1,-1)--(2,-0.5);
	
	\draw[fill=black!30!white] (0,0) circle (.25);	
	
	\node at (1.5, 1.6) {\color{red} $\tilde{\chi}^0$};
	\node at (1.5, -1.6) {\color{red} $\tilde{\chi}^0$};
	\node at (0.4, 0.8) {\color{red} $\tilde{t}$};
	\node at (0.4, -0.8) {\color{red} $\tilde{t}$};	
	\node at (2.2, 0.4) {$\bar{t}$};
	\node at (2.2, -0.4) {$\bar{t}$};

  	\end{scope}
	
 \end{tikzpicture}
\end{center}
\caption{Leading order Feynman diagram for the compressed gluino scenario (left) and direct stop pair production (right).  Note that since the gluino is its own anti-particle, the two on-shell stops can have opposite or the same charge.}
\label{fig:feynman}
\end{figure}
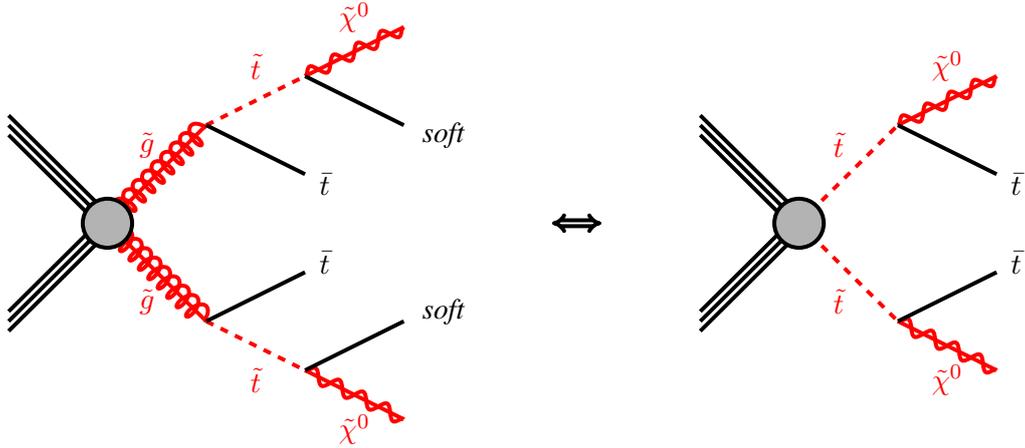

When the mass splitting between the stop and the lightest neutralino is very small, many of the traditional techniques for identifying gluino pair production are ineffective.  Recognizing the phenomenological similarity between $\tilde{g}\rightarrow t\tilde{t}$ when $m_{\tilde{t}}\sim m_{\tilde{\chi}_1^0}$ and the direct production of $\tilde{t}\rightarrow t\tilde{\chi}_1^0$ when $m_{\tilde{t}}\gg m_{\tilde{\chi}_1^0}$, depicted in Fig.~\ref{fig:feynman}, suggests that searches for the later can be recast as searches for gluinos.   Appendices~\ref{sec:compressed} and~\ref{sec:compressed2} systematically describes the various compressed scenarios possible in gluino-mediated stop/sbottom production, but the remainder of this paper focuses on the {\it gluino mediated compressed stop} scenario.

\section{Limits on Compressed Gluino Mediated Stop Production}
\label{sec:limits}

The lost sensitivity to compressed $\tilde{g}\rightarrow t\tilde{t}$ from direct gluino searches with multi-top quark, multi-$b$ quark, or multi-lepton final states can be recovered by direct stop searches.  As shown schematically in Fig.~\ref{fig:feynman}, the final state for the gluino mediated compressed stop is the same as the direct stop production.  There are only subtle differences due to the fact that the gluino is a fermionic color octet, instead of a scalar triplet like the stop, so there will be small differences in angular distributions and radiation patterns between jets.  However, most analysis techniques are not sensitive to these effects.  One non-negligible difference is the electric charge, as stops can have the same charge when from gluinos, but must be oppositely charged for direct stop production.  For this reason, gluino searches with multi-top/$b$/lepton final states loose sensitivity to the gluino mediated compressed stop scenario, but same-sign lepton searches can retain sensitivity.  However, we will soon see that one- and zero-lepton searches will be more powerful, due to the much larger branching ratio.  Both ATLAS~\cite{Aad:2012ywa,Aad:2012xqa,Aad:2012uu,Aad:2014kra,Aad:2014bva,Aad:2014qaa,Aad:2014mfk,Aad:2014nra} and CMS~\cite{Chatrchyan:2013xna,Khachatryan:2015wza,CMS-PAS-SUS-13-009,Khachatryan:2014doa,Chatrchyan:2014goa,Chatrchyan:2012uea} have searched for direct stop pair production using a variety of techniques to target the rich phenomenology possible in stop decays. The current limits on direct stop production for a massless neutralino reach about $m_{\tilde{t}}\sim 700$ GeV in both the one lepton and zero lepton final states.  In the next section, we will discuss how to recast these limits for gluino pair production.

\subsection{Reinterpreting Direct Stop Limits}

For a massless neutralino, one can translate the limits on\footnote{To ease the notation, we will drop the subscripts on the stop and the neutralino - the smallest mass eigenstates are assumed throughout.} $\tilde{t}_1\rightarrow t\tilde{\chi}_1^0$ into limits on $\tilde{g}\rightarrow t\tilde{t}$ with $m_{\tilde{t}}^{\tilde{g}}\sim m_{\tilde{\chi}^0}^{\tilde{g}}\sim 0$ by solving the equation $\sigma_{\tilde{t}}(m_{\tilde{t}})=\sigma_{\tilde{g}}(m_{\tilde{g}}^{\tilde{g}})$, from e.g. the numbers published by Ref.~\cite{Kramer:2012bx}.  The superscript $\tilde{g}$ is used to distinguish the mass hierarchy in the direct stop production (no subscript) from the mass hierarchy in the gluino mediated stop production in the reinterpretation.  To extend the limits toward $m_{\tilde{t}}^{\tilde{g}}>0$, one needs to choose stop masses $m_{\tilde{t}}^{\tilde{g}}$ such that the top quark and neutralino $p_T$ from the gluino decay at $m_{\tilde{g}}^{\tilde{g}}$ should be comparable to the top quark and neutralino $p_T$ from the stop decay at $m_{\tilde{t}}$.  Define the two-body phase space momentum:

$$p(M,m)=\sqrt{\frac{\left(M^2-(m_t-m)^2\right)\left(M^2-(m_t+m)^2\right)}{4M^2}},$$

\noindent where $m_t\sim 175$ GeV is the top quark mass~\cite{Agashe:2014kda}.  Then, given a pair $(m_{\tilde{t}},m_{\tilde{\chi}^0})$ excluded by a direct stop search, and $m_{\tilde{\chi}^0}^{\tilde{g}}$ from $\sigma_{\tilde{t}}(m_{\tilde{t}})=\sigma_{\tilde{g}}(m_{\tilde{g}}^{\tilde{g}})$, the re-interpreted excluded stop mass $m_{\tilde{t}}^{\tilde{g}}$ is given by the solution to $p(m_{\tilde{t}},m_{\tilde{\chi}^0})=p(m_{\tilde{g}}^{\tilde{g}},m_{\tilde{t}}^{\tilde{g}})$.  The solution is quartic in $m_{\tilde{t}_1}^{\tilde{g}}$, so in general there can be up to four real solutions.  Fortunately, two solutions are negative (or imaginary) and of the two possible positive solutions, only one can be smaller than $m_{\tilde{g}}^{\tilde{g}}$ and thus there is at most one physical solution.  The translation for the high mass stops given in the recent 8 TeV ATLAS stop search\footnote{Similar limits exist for the ATLAS all-hadronic search~\cite{Aad:2014bva} and both the CMS leptonic~\cite{Chatrchyan:2013xna} and hadronic searches~\cite{Khachatryan:2015wza}.} in the one lepton final state~\cite{Aad:2014kra} are shown in Table~\ref{tab:translate}.

\begin{table}[h]
\begin{center}
\begin{tabular}{|c|c|c|c|c|}
\hline
$m_{\tilde{t}}$ & $\max m_{\tilde{\chi}_1^0}$ & $\sigma_{\tilde{t}}(m_{\tilde{t}})$ & $m_{\tilde{g}}^{\tilde{g}}$ & $\max m_{\tilde{t}}^{\tilde{g}}$ \\
\hline
675             & 100                         & 0.011                               & 1090                                            & 670                                             \\
625             & 220                         & 0.018                               & 1030                                            & 690                                             \\
600             & 240                         & 0.025                               & 995                                             & 680                                             \\
550             & 240                         & 0.045                               & 930                                             & 660      \\\hline                                    
\end{tabular}
\caption{The corresponds between the limits on stop mass from the direct search presented in Ref.~\cite{Aad:2014kra}.  All masses are given in GeV and all cross sections are in pb.  The notation $\max X$ means the maximum value of $X$ at the given $m_{\tilde{t}}$ or $m_{\tilde{g}}^{\tilde{g}}$.  The first two columns are extracted from the (observed) exclusion plots in Ref.~\cite{Aad:2014kra}, the third column is from Ref.~\cite{Kramer:2012bx}, the fourth column uses Ref.~\cite{Kramer:2012bx} to solve $\sigma_{\tilde{t}}(m_{\tilde{t}})=\sigma_{\tilde{g}}(m_{\tilde{g}}^{\tilde{g}})$ and the last column is derived using the other columns and the two-body phase space equation given in the text.}
\label{tab:translate}
\end{center}
\end{table}

The above procedure will produce limits for the gluino mediated compressed stop so long as $p(m_{\tilde{g}}^{\tilde{g}},m_{\tilde{t}}^{\tilde{g}}) < p(m_{\tilde{t}},0)$.  However, this is an artificial constraint - exclusion power increases in the top/neutralino momentum $p(m_{\tilde{t}},x)$, which is increasing for decreasing $x$.  Therefore, it is a safe assumption that if the point $p(m_{\tilde{g}}^{\tilde{g}},m_{\tilde{t}}^{\tilde{g}})$ is excluded, then the point $p(m_{\tilde{g}}^{\tilde{g}},0)$ will also be excluded.  One can take this argument one step further.  If the maximum stop mass excluded by direct searches is $m_{\tilde{t}}^\text{max}$, then it is artificial for the maximum excluded gluino mass to be $\sigma_{\tilde{t}}(m_{\tilde{t}})=\sigma_{\tilde{g}}(m_{\tilde{g}}^{\tilde{g}})$, because the acceptance increases with gluino mass.  One can extend the limits to larger gluino masses by noting that the acceptance $\mathcal{A}$ for a particular model to pass all signal region requirements depends only on $p(m_{\tilde{t}},m_{\tilde{\chi}^0})$ and not on $m_{\tilde{t}}$ or $m_{\tilde{\chi}^0}$ directly and then extrapolating to higher values of $p$ beyond $p(m_{\tilde{t}}^\text{max},0)$.  Figure~\ref{fig:excl} shows the acceptance from the ATLAS stop search in the one lepton final state~\cite{Aad:2014kra} as a function of $p$.  The observation that the acceptance only depends on $p$ is confirmed by the fact that there is one curve independent of the stop mass.  For large values of $p$, the acceptance should be roughly linear in $p$ as the missing momentum in the event is linear in $p$.  Therefore, we fit the curve beyond $p=200$ to a straight line for extrapolating the acceptance to higher values of $p$.  Values of $(m_{\tilde{g}}^{\tilde{g}},m_{\tilde{t}}^{\tilde{g}})$ can be declared excluded if $\sigma_{\tilde{g}}(m_{\tilde{g}}^{\tilde{g}})\times \mathcal{A}(p(m_{\tilde{g}}^{\tilde{g}},m_{\tilde{t}}^{\tilde{g}})) > n_\text{excluded}$, where $n_\text{excluded}$ is the excluded beyond the Standard Model number of events from the published search (5.3 events).

One can do even better than naively recasting limits based on $n_\text{excluded}$ by tightening thresholds on the key variables $E_\text{T}^\text{miss}$, $m_\text{T}$, $(a)m_\text{T2}$, $m^\text{jet}$, etc.~\cite{Bai:2012gs,Aad:2014kra,Graesser:2012qy,Nachman:2013bia}, but studying this change would require a careful assessment of the change in the background yield which is beyond the scope of this paper.  Limits can additionally be improved by combining multiple channel (one lepton and zero lepton).

 \begin{figure}[h!]
 \begin{center}
 \includegraphics[width=0.8\textwidth]{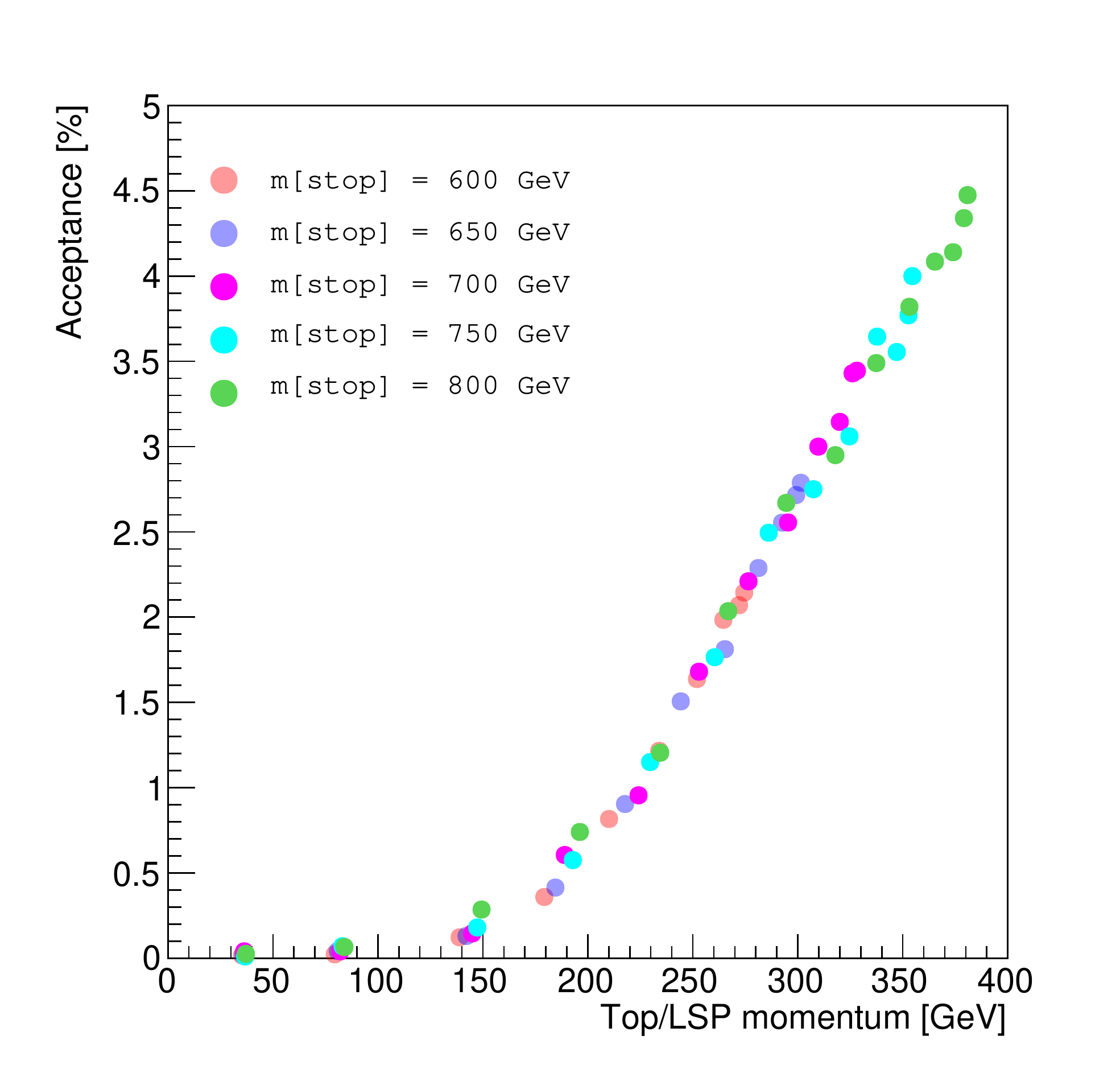}
 \caption{The acceptance from Aux Fig. 25 in Ref.~\cite{aux2}.  The acceptance is defined as the fraction of simulated signal events that pass a particle-level version of the analysis (no detector simulation).  Note that the ratio of the particle-level acceptance to the acceptance using the full detector simulation (Aux Fig. 29 in Ref.~\cite{aux2}) are all within a few percent of one in the relevant region and so can be safely ignored.  In the region beyond 600 GeV, a straight line is fit with parameters $\mathcal{A} = 0.02\frac{\%}{\text{GeV}}\times p  -3.8\%$.}
 \label{fig:excl}
  \end{center}
 \end{figure}

\clearpage

\subsection{Derived Limits}

The re-casted direct stop limits are shown Fig.~\ref{fig:excl} alongside existing limits from the ATLAS same-sign search~\cite{Aad:2014pda} and the inclusive one lepton\footnote{A similar search exists in the zero lepton final state, with slightly weaker limits~\cite{Aad:2014wea} } search~\cite{Aad:2015mia}.   The same-sign limits are optimistic because the selection in Ref.~\cite{Aad:2014pda} requires a third hard jet, which is not part of the leading order description of the final state.  Estimates based on calculations with MadGraph5\_aMC@NLO version 5.2.1.1~\cite{Alwall:2014hca} indicate that the fraction of the time an additional jet from initial or final state radiation has enough $p_T$ to pass the jet selection is roughly 40\%.  This agrees well with the three jet selection efficiency published in auxiliary material Table 64~\cite{aux} of the ATLAS search for a model with large top mass for which kinematically the soft $c$-quark jets will not pass the hard jet $p_T$ threshold.  As the mass splitting between the stop and the neturalino goes to zero, the reduction in the limit for the highest mass splitting reduces by $\lesssim100$~GeV (not shown).  The inclusive one lepton search is based on generic variables such as $E_\text{T}^\text{miss}$, $m_\text{T}$, effective mass, etc. and is not optimized for the $t\bar{t}+E_\text{T}^\text{miss}$ final state (the limits may even degrade as $m_{\tilde{t}}\rightarrow m_{\tilde{\chi}^0}$).  The improvement over these existing analyses for the reinterpreted direct search are shown in shaded blue in Fig.~\ref{fig:excl}.  The darkest blue is from the strict re-interpretation based on the strategy leading up to Table~\ref{tab:translate}.  The light blue area below the dark blue area is assumed excluded because the signal efficiency increases for the larger mass splitting.  The light blue area to the right of the dashed line is from interpolating and extrapolating the efficiency and comparing to the published limit on allowed beyond the Standard Model events.  For a 1.1 TeV gluino, the inclusive one lepton limit is extended vertically by about 300 GeV.

 \begin{figure}[h!]
 \begin{center}
 \includegraphics[width=0.65\textwidth]{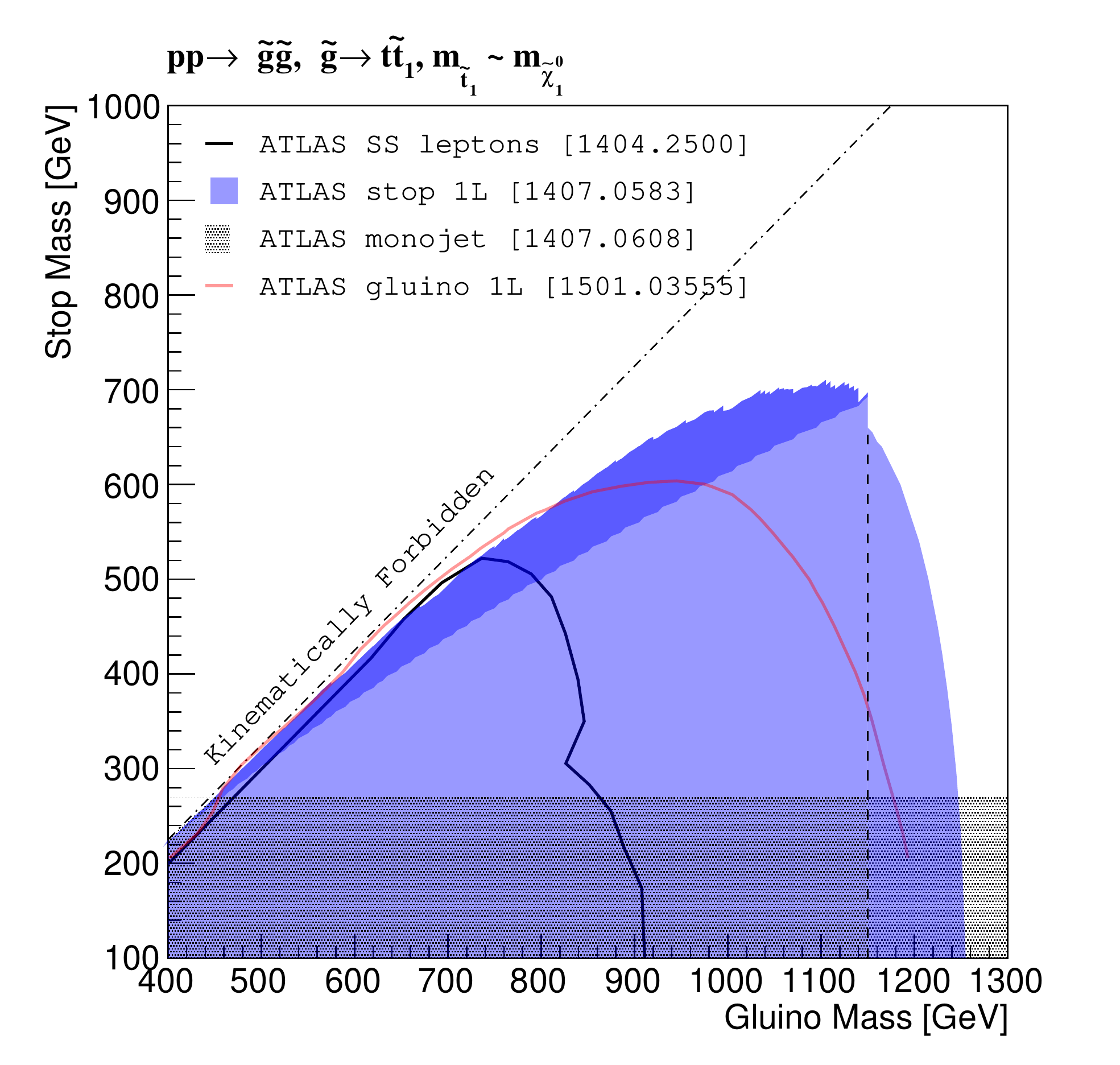}
 \caption{A comparison of existing limits and the re-interpretation of the ATLAS direct stop search in the one lepton channel at 8 TeV.  The expected limits (based on the $\text{CL}_{s}$ procedure~\cite{Read:2002hq}) are used to control for statistical fluctuations in the observations.  The hatched region is from the 8 TeV ATLAS search for compressed direct stop production via an ISR monojet~\cite{Aad:2014nra}.  The red line is from the ATLAS inclusive one lepton search~\cite{Aad:2015mia} (Fig. 18a) and the black line is from the ATLAS same-sign lepton search~\cite{Aad:2014pda}.  The shaded region is from the re-interpreted direct stop search, as described in the text.}
 \label{fig:excl}
  \end{center}
 \end{figure}

\clearpage
\newpage

\section{Conclusions and Future Outlook}
\label{sec:conc}

Natural, weak scale SUSY is a compelling paradigm for models of new physics.  The sensitivity to such models with a nearly mass degenerate stop and neutralino can be extended by repurposing searches for direct stop pair production.  The major implications of this work are:

\begin{itemize}

\item Even though the decay products of a stop (or a sbottom) might be missed due to detector thresholds, the other gluino decay product(s) can empower complementary search techniques targeting one step decay chains with fewer objects in the final state.  Direct searches could do even better with targeted optimization to higher $p_T$ final states.  No limits currently exist (that the author knows of) for the gluino mediated compressed sbottom.\vspace{3mm}
\item At 13 TeV, searches for direct stop/sbottom pair production might be able to discover SUSY much earlier than expected because the direct stop/sbottom cross section is much smaller than the gluino cross section.  Table~\ref{tab:13TeV} summarizes the relative increase in cross sections.  Larger masses generally have a larger increase in cross section from 8 to 13 TeV center of mass energy.  Thus, since the sensitivity to gluino masses at 8 TeV is much higher than the stop masses, the increase from 8 to 13 TeV is bigger for the gluinos.  At the edge of the 8 TeV sensitivity, the expected increase in the yield of gluinos is twice the corresponding yield for stops.
\end{itemize}

\noindent The discovery of SUSY could be within reach of the Run II of the LHC.  All possibilities for natural SUSY should be targeted, including those with compressed scenarios.  If there is a light enough gluino to mediate, more territory for light stops and sbottoms will be accessible to the direct searches with the early data.

\begin{table}[h!]
\begin{center}
\begin{tabular}{|c|c|c|c|c|c|c|}
\hline
$m_{\tilde{t}}$ & $\sigma_{\tilde{t}}^\text{8 TeV}(m_{\tilde{t}})$ & $m_{\tilde{g}}^{\tilde{g}}$ & $\sigma_{\tilde{t}}^\text{13 TeV}(m_{\tilde{t}})$ &  $\sigma^\text{13 TeV}_{\tilde{g}}(m_{\tilde{g}}^{\tilde{g}})$ & $\sigma_{\tilde{t}}(m_{\tilde{t}})$ 13 TeV/8 TeV&  $\sigma_{\tilde{g}}(m_{\tilde{g}})$ 13 TeV/8 TeV \\
\hline
600             & 0.03                                      & 1000                        & 0.2                                   & 0.3                                               & 7.0                  & 13.6        \\
700             & 0.008                                     & 1125                        & 0.07                                  & 0.1                                               & 8.3                  & 17.1        \\
800             & 0.003                                     & 1250                        & 0.03                                 & 0.06                                             & 9.8                  & 21.7    \\
\hline   
\end{tabular}
\label{tab:13TeV}
\caption{The expected increase in yields for the direct stop search and the re-interpreted gluino search from 8 to 13 TeV.  The first column is the stop mass in GeV, the second column is the stop cross section at 8 TeV in pb from Ref.~\cite{Kramer:2012bx}.  The third column is in pb and also uses Ref.~\cite{Kramer:2012bx} to solve $\sigma_{\tilde{t}_1}(m_{\tilde{t}_1})=\sigma_{\tilde{g}}(m_{\tilde{g}}^{\tilde{g}})$. The fourth and fifth columns give the cross sections for stop and gluino production at 13 TeV from Ref.~\cite{Borschensky:2014cia}.  The last two columns give the ratio of the increase in yields for stop and gluino production, respectively. }
\end{center}
\end{table}

\section{Acknowledgements}

We would like to thank Till Eifert and Michael Peskin for useful discussions and Tommaso Lari for referring us to the limits in Ref.~\cite{Aad:2015mia}.  BN is supported by the NSF Graduate Research Fellowship under Grant No. DGE-4747 and by the Stanford Graduate Fellowship.

\clearpage
\newpage

\appendix

\section{Compressed Scenarios}

When $m_{\tilde{g}} > m_t+m_{\tilde{t}}$ or $m_{\tilde{t}}\gg m_{\tilde{g}}$, gluinos can decay via an on- or off-shell stop to a $t\bar{t}\tilde{\chi}_1^0$ final state.  Since each top quark decays to a $b$ quark and an on-shell $W$ boson, the $t\bar{t}\chi_1^0$ final state results in four high $p_T$ $b$-quarks and large missing energy from the neutralinos ($\tilde{\chi}_1^0$) with the possibility of many leptons from the $W$ decays, some pairs of which can have the same charge.   A natural SUSY spectrum would also suggest that the SUSY partner of the left-handed bottom quark, the sbottom ($\tilde{b}$), should also be light and if produced via $\tilde{g}\rightarrow b\tilde{b}$ can give rise to four $b$-quarks and up to four leptons as well if $\tilde{b}\rightarrow t\chi_1^\pm$.  There are not many Standard Model processes that produce many high energy $b$-quarks, multi- or same-sign leptons in association with large missing energy and so such techniques are very powerful.   

However, the high energy, high multiplicity final states from gluino pair production may become difficult to identify experimentally for compressed mass spectra.  In this section, we systematically consider such scenarios by enumerating all mass hierarchies.  Throughout, $\tilde{g}\rightarrow t\tilde{t}$ (or $\tilde{g}\rightarrow b\tilde{b}$) and $m_{\tilde{t}}<m_{\tilde{g}}$ (or $m_{\tilde{b}}<m_{\tilde{g}}$).   Stops and sbottoms have both flavor conserving and flavor changing decays.   The flavor preserving decays are discussed in Sec.~\ref{sec:compressed} and those with flavor changing decays are discussed in Sec.~\ref{sec:compressed2}.

\subsection{Flavor Preserving Decays}
\label{sec:compressed}

\begin{figure}[h!]

\begin{center}
\begin{tikzpicture}[line width=1.5 pt, scale=1.3]
	
	\node at (0, 4) {$m_{\tilde{g}}$};
	\node at (0, 2) {$m_{\tilde{q}}$};
	\node at (0, 0) {$m_{\tilde{\chi}_1^0}$};	

	\draw[dotted,color=gray!40!white] (0.5,4)--(8.5,4);		
	\draw (1,4)--(2,4);	
	\draw (3,4)--(4,4);	
	\draw (5,4)--(6,4);	
	\draw (7,4)--(8,4);		

	\draw[dotted,color=gray!40!white] (0.5,2)--(2.25,2);
	\draw[dotted,color=gray!40!white] (2.25,2)--(2.75,3.5);	
	\draw[dotted,color=gray!40!white] (2.75,3.5)--(4.25,3.5);		
	\draw[dotted,color=gray!40!white] (4.25,3.5)--(4.75,2);	
	\draw[dotted,color=gray!40!white] (4.75,2)--(6.25,2);	
	\draw[dotted,color=gray!40!white] (6.25,2)--(6.75,3.5);	
	\draw[dotted,color=gray!40!white] (6.75,3.5)--(8.5,3.5);						
	\draw (1,2)--(2,2);	
	\draw (3,3.5)--(4,3.5);	
	\draw (5,2)--(6,2);	
	\draw (7,3.5)--(8,3.5);	

	\draw[dotted,color=gray!40!white] (0.5,0)--(4.25,0);
	\draw[dotted,color=gray!40!white] (4.25,0)--(4.75,1.5);	
	\draw[dotted,color=gray!40!white] (4.75,1.5)--(6.25,1.5);		
	\draw[dotted,color=gray!40!white] (6.25,1.5)--(6.75,3);	
	\draw[dotted,color=gray!40!white] (6.75,3.0)--(8.5,3.);						
	\draw (1,0)--(2,0);	
	\draw (3,0)--(4,0);	
	\draw (5,1.5)--(6,1.5);	
	\draw (7,3.)--(8,3.);	

	\node at (1.5, -0.5) {(a)};
	\node at (3.5, -0.5) {(b)};	
	\node at (5.5, -0.5) {(c)};
	\node at (7.5, -0.5) {(d)};	
	
 \end{tikzpicture}
\end{center}
\caption{The possible mass hierarchies for $\tilde{g}\rightarrow q\tilde{q}$, $q=t$ or $b$.  The scenarios (a) and (b) are well-covered by traditional gluino searches.  Direct gluino searches may loose sensitivity in scenario (c), but this loss may be recovered by direct stop/sbottom searches.  Sensitivity to scenario (d) requires additional radiation in the event that makes it challenging for direct SUSY searches.}
\label{fig:masshierarchy}
\end{figure}
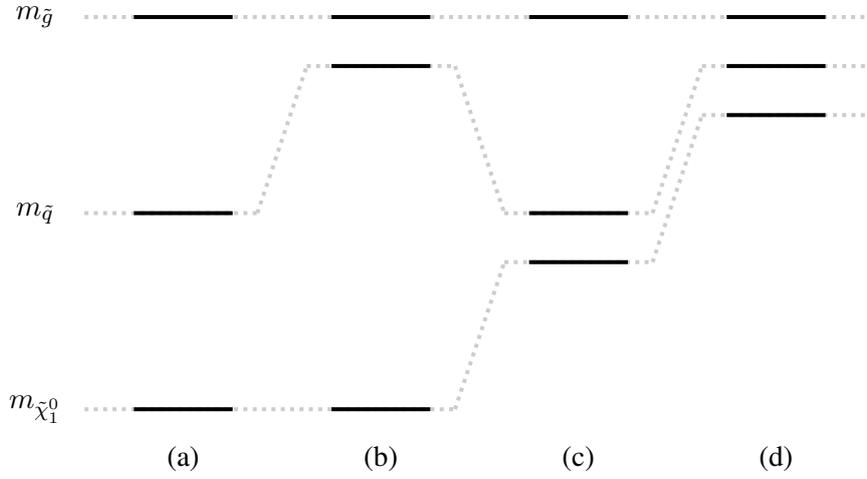

First, suppose that $\tilde{t}\rightarrow t\chi_1^0$ or $\tilde{b}\rightarrow b\chi_1^0$.  The possible hierarchies are depicted schematically in Fig.~\ref{fig:masshierarchy}.  The large mass splittings in Fig.~\ref{fig:masshierarchy}a are well covered by traditional gluino searches that look for e.g. multi-$b$ final states with large missing momentum.  The highly compressed spectrum in Fig.~\ref{fig:masshierarchy}d is not possible to identify without additional initial or final state radiation that would lead to a mono-jet topology, which is beyond the scope of this analysis.  For the $\tilde{g}\rightarrow t\tilde{t}$, the scenario shown in Fig.~\ref{fig:masshierarchy}b cannot be too compressed because the top quark would have to be very off-shell for its decay products to be too soft to detect.  As $m_{\tilde{g}}-m_{\tilde{t}}$ falls below $m_t$, the top and stop resonances will go off-shell and compete in order to conserve energy.  Since $m_{\tilde{t}}\gg m_{\chi_1^0}$, the stop width is unsuppressed.  When the off-shell top mass is below the $W$ mass, the price for taking the top further off-shell is more than for the stop and so it is likely that the scenario for stop production in Fig.~\ref{fig:masshierarchy}b is still well-covered by existing searches.  In contrast, for $\tilde{g}\rightarrow b\tilde{b}$, the scenario shown in Fig.~\ref{fig:masshierarchy}b can be compressed all the way to $m_{\tilde{g}}-m_{\tilde{b}}\sim m_b$ in which case the $b$-quark from the direct gluino decay might be too soft to measure experimentally.  This is phenomenologically similar to the models in Fig.~\ref{fig:masshierarchy}c.  The scenarios shown in Fig.~\ref{fig:masshierarchy}c are a clear source of compressed gluino decays for which traditional searches may loose sensitivity and are well-motivated by stop-neutralino co-annihilation to produce the correct DM cosmological abundance~\cite{deSimone:2014pda,Delgado:2012eu,Profumo:2004at}.  When the stop (or sbottom) and neutralino are nearly mass degenerate, the phenomenology of the $\tilde{g}\rightarrow t\tilde{t}$ is basically the same as $\tilde{g}\leadsto t\tilde{\chi}_1^0$ (or replace $t\leftrightarrow b$ and $\tilde{t}\leftrightarrow\tilde{b}$), and are covered by the analyses discussed in the main body of the paper.

\subsection{Flavor Changing Decays}
\label{sec:compressed2}

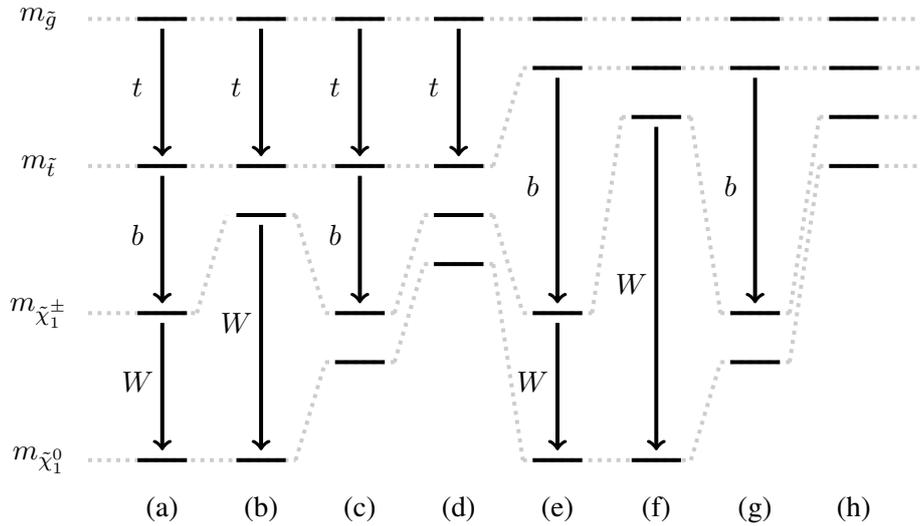
\begin{figure}[h!]

\begin{center}
\begin{tikzpicture}[line width=1.5 pt, scale=1.3]
	
	\node at (-0.5, 4.5) {$m_{\tilde{g}}$};
	\node at (-0.5, 3) {$m_{\tilde{t}}$};
	\node at (-0.5, 1.5) {$m_{\tilde{\chi}_1^\pm}$};
	\node at (-0.5, 0) {$m_{\tilde{\chi}_1^0}$};	

	\draw[dotted,color=gray!40!white] (0.,4.5)--(8.5,4.5);		
	\draw (0.5,4.5)--(1.,4.5);	
	\draw (1.5,4.5)--(2.,4.5);	
	\draw (2.5,4.5)--(3.,4.5);	
	\draw (3.5,4.5)--(4.,4.5);	
	\draw (4.5,4.5)--(5.,4.5);	
	\draw (5.5,4.5)--(6.,4.5);	
	\draw (6.5,4.5)--(7.,4.5);	
	\draw (7.5,4.5)--(8.,4.5);		
	
	\draw[dotted,color=gray!40!white] (0.,3)--(4.1,3);
	\draw[dotted,color=gray!40!white] (4.1,3)--(4.4,4);
	\draw[dotted,color=gray!40!white] (4.4,4)--(8.5,4);		
	\draw (0.5,3)--(1.,3);	
	\draw (1.5,3)--(2.,3);	
	\draw (2.5,3)--(3.,3);	
	\draw (3.5,3)--(4.,3);	
	\draw (4.5,4)--(5.,4);	
	\draw (5.5,4)--(6.,4);	
	\draw (6.5,4)--(7.,4);	
	\draw (7.5,4)--(8.,4);	
	
	\draw[dotted,color=gray!40!white] (0.,1.5)--(1.1,1.5);	
	\draw[dotted,color=gray!40!white] (1.1,1.5)--(1.4,2.5);
	\draw[dotted,color=gray!40!white] (1.5,2.5)--(2.1,2.5);
	\draw[dotted,color=gray!40!white] (2.1,2.5)--(2.4,1.5);
	\draw[dotted,color=gray!40!white] (2.4,1.5)--(3.1,1.5);
	\draw[dotted,color=gray!40!white] (3.1,1.5)--(3.4,2.5);
	\draw[dotted,color=gray!40!white] (3.4,2.5)--(4.1,2.5);
	\draw[dotted,color=gray!40!white] (4.1,2.5)--(4.4,1.5);
	\draw[dotted,color=gray!40!white] (4.4,1.5)--(5.1,1.5);
	\draw[dotted,color=gray!40!white] (5.1,1.5)--(5.4,3.5);
	\draw[dotted,color=gray!40!white] (5.4,3.5)--(6.1,3.5);
	\draw[dotted,color=gray!40!white] (6.1,3.5)--(6.4,1.5);
	\draw[dotted,color=gray!40!white] (6.4,1.5)--(7.1,1.5);
	\draw[dotted,color=gray!40!white] (7.1,1.5)--(7.4,3.5);
	\draw[dotted,color=gray!40!white] (7.4,3.5)--(8.5,3.5);		
	\draw (0.5,1.5)--(1.,1.5);	
	\draw (1.5,2.5)--(2.,2.5);	
	\draw (2.5,1.5)--(3.,1.5);	
	\draw (3.5,2.5)--(4.,2.5);	
	\draw (4.5,1.5)--(5.,1.5);	
	\draw (5.5,3.5)--(6.,3.5);	
	\draw (6.5,1.5)--(7.,1.5);	
	\draw (7.5,3.5)--(8.,3.5);	

	\draw[dotted,color=gray!40!white] (0.,0)--(2.1,0);
	\draw[dotted,color=gray!40!white] (2.1,0)--(2.4,1);
	\draw[dotted,color=gray!40!white] (2.4,1)--(3.1,1);
	\draw[dotted,color=gray!40!white] (3.1,1)--(3.4,2);	
	\draw[dotted,color=gray!40!white] (3.4,2)--(4.1,2);	
	\draw[dotted,color=gray!40!white] (4.1,2)--(4.4,0);
	\draw[dotted,color=gray!40!white] (4.4,0)--(6.1,0);
	\draw[dotted,color=gray!40!white] (6.1,0)--(6.4,1);	
	\draw[dotted,color=gray!40!white] (6.4,1)--(7.1,1);	
	\draw[dotted,color=gray!40!white] (7.1,1)--(7.4,3);
	\draw[dotted,color=gray!40!white] (7.4,3)--(8.5,3);		
	\draw (0.5,0)--(1.,0);	
	\draw (1.5,0)--(2.,0);	
	\draw (2.5,1.)--(3.,1.);	
	\draw (3.5,2.)--(4.,2);	
	\draw (4.5,0)--(5.,0);	
	\draw (5.5,0)--(6.,0);	
	\draw (6.5,1)--(7.,1);	
	\draw (7.5,3)--(8.,3);	

	\draw[->] (0.75,4.4)--(0.75,3.1);	
	\node at (0.5, 3.8) {$t$};

	\draw[->] (1.75,4.4)--(1.75,3.1);	
	\node at (1.5, 3.8) {$t$};

	\draw[->] (2.75,4.4)--(2.75,3.1);	
	\node at (2.5, 3.8) {$t$};

	\draw[->] (3.75,4.4)--(3.75,3.1);	
	\node at (3.5, 3.8) {$t$};

	\draw[->] (0.75,2.9)--(0.75,1.6);	
	\node at (0.5, 2.3) {$b$};

	\draw[->] (2.75,2.9)--(2.75,1.6);	
	\node at (2.5, 2.3) {$b$};

	\draw[->] (4.75,3.9)--(4.75,1.6);	
	\node at (4.5, 2.8) {$b$};

	\draw[->] (6.75,3.9)--(6.75,1.6);	
	\node at (6.5, 2.8) {$b$};

	\draw[->] (0.75,1.4)--(0.75,0.1);	
	\node at (0.5,0.8) {$W$};

	\draw[->] (1.75,2.4)--(1.75,0.1);	
	\node at (1.5,1.4) {$W$};

	\draw[->] (4.75,1.4)--(4.75,0.1);	
	\node at (4.5,0.8) {$W$};

	\draw[->] (5.75,3.4)--(5.75,0.1);	
	\node at (5.5,1.8) {$W$};

	\node at (0.75, -0.5) {(a)};
	\node at (1.75, -0.5) {(b)};	
	\node at (2.75, -0.5) {(c)};
	\node at (3.75, -0.5) {(d)};	
	\node at (4.75, -0.5) {(e)};	
	\node at (5.75, -0.5) {(f)};		
	\node at (6.75, -0.5) {(g)};		
	\node at (7.75, -0.5) {(h)};		
	
 \end{tikzpicture}
\end{center}
\caption{The possible mass hierarchies for $\tilde{g}\rightarrow t\tilde{t}$, $\tilde{t}\rightarrow b\chi_1^\pm$.  The arrows denote which, if any, high energy particles are produced in the cascade decay.}
\label{fig:masshierarchy3}
\end{figure}

\begin{figure}[h!]

\begin{center}
\begin{tikzpicture}[line width=1.5 pt, scale=1.3]
	
	\node at (-0.5, 4.5) {$m_{\tilde{g}}$};
	\node at (-0.5, 3) {$m_{\tilde{b}}$};
	\node at (-0.5, 1.5) {$m_{\tilde{\chi}_1^\pm}$};
	\node at (-0.5, 0) {$m_{\tilde{\chi}_1^0}$};	

	\draw[dotted,color=gray!40!white] (0.,4.5)--(8.5,4.5);		
	\draw (0.5,4.5)--(1.,4.5);	
	\draw (1.5,4.5)--(2.,4.5);	
	\draw (2.5,4.5)--(3.,4.5);	
	\draw (3.5,4.5)--(4.,4.5);	
	\draw (4.5,4.5)--(5.,4.5);	
	\draw (5.5,4.5)--(6.,4.5);	
	\draw (6.5,4.5)--(7.,4.5);	
	\draw (7.5,4.5)--(8.,4.5);		
	
	\draw[dotted,color=gray!40!white] (0.,3)--(4.1,3);
	\draw[dotted,color=gray!40!white] (4.1,3)--(4.4,4);
	\draw[dotted,color=gray!40!white] (4.4,4)--(8.5,4);		
	\draw (0.5,3)--(1.,3);	
	\draw (1.5,3)--(2.,3);	
	\draw (2.5,3)--(3.,3);	
	\draw (3.5,3)--(4.,3);	
	\draw (4.5,4)--(5.,4);	
	\draw (5.5,4)--(6.,4);	
	\draw (6.5,4)--(7.,4);	
	\draw (7.5,4)--(8.,4);	
	
	\draw[dotted,color=gray!40!white] (0.,1.5)--(1.1,1.5);	
	\draw[dotted,color=gray!40!white] (1.1,1.5)--(1.4,2.5);
	\draw[dotted,color=gray!40!white] (1.5,2.5)--(2.1,2.5);
	\draw[dotted,color=gray!40!white] (2.1,2.5)--(2.4,1.5);
	\draw[dotted,color=gray!40!white] (2.4,1.5)--(3.1,1.5);
	\draw[dotted,color=gray!40!white] (3.1,1.5)--(3.4,2.5);
	\draw[dotted,color=gray!40!white] (3.4,2.5)--(4.1,2.5);
	\draw[dotted,color=gray!40!white] (4.1,2.5)--(4.4,1.5);
	\draw[dotted,color=gray!40!white] (4.4,1.5)--(5.1,1.5);
	\draw[dotted,color=gray!40!white] (5.1,1.5)--(5.4,3.5);
	\draw[dotted,color=gray!40!white] (5.4,3.5)--(6.1,3.5);
	\draw[dotted,color=gray!40!white] (6.1,3.5)--(6.4,1.5);
	\draw[dotted,color=gray!40!white] (6.4,1.5)--(7.1,1.5);
	\draw[dotted,color=gray!40!white] (7.1,1.5)--(7.4,3.5);
	\draw[dotted,color=gray!40!white] (7.4,3.5)--(8.5,3.5);		
	\draw (0.5,1.5)--(1.,1.5);	
	\draw (1.5,2.5)--(2.,2.5);	
	\draw (2.5,1.5)--(3.,1.5);	
	\draw (3.5,2.5)--(4.,2.5);	
	\draw (4.5,1.5)--(5.,1.5);	
	\draw (5.5,3.5)--(6.,3.5);	
	\draw (6.5,1.5)--(7.,1.5);	
	\draw (7.5,3.5)--(8.,3.5);	

	\draw[dotted,color=gray!40!white] (0.,0)--(2.1,0);
	\draw[dotted,color=gray!40!white] (2.1,0)--(2.4,1);
	\draw[dotted,color=gray!40!white] (2.4,1)--(3.1,1);
	\draw[dotted,color=gray!40!white] (3.1,1)--(3.4,2);	
	\draw[dotted,color=gray!40!white] (3.4,2)--(4.1,2);	
	\draw[dotted,color=gray!40!white] (4.1,2)--(4.4,0);
	\draw[dotted,color=gray!40!white] (4.4,0)--(6.1,0);
	\draw[dotted,color=gray!40!white] (6.1,0)--(6.4,1);	
	\draw[dotted,color=gray!40!white] (6.4,1)--(7.1,1);	
	\draw[dotted,color=gray!40!white] (7.1,1)--(7.4,3);
	\draw[dotted,color=gray!40!white] (7.4,3)--(8.5,3);		
	\draw (0.5,0)--(1.,0);	
	\draw (1.5,0)--(2.,0);	
	\draw (2.5,1.)--(3.,1.);	
	\draw (3.5,2.)--(4.,2);	
	\draw (4.5,0)--(5.,0);	
	\draw (5.5,0)--(6.,0);	
	\draw (6.5,1)--(7.,1);	
	\draw (7.5,3)--(8.,3);	

	\draw[->] (0.75,4.4)--(0.75,3.1);	
	\node at (0.5, 3.8) {$b$};

	\draw[->] (1.75,4.4)--(1.75,3.1);	
	\node at (1.5, 3.8) {$b$};

	\draw[->] (2.75,4.4)--(2.75,3.1);	
	\node at (2.5, 3.8) {$b$};

	\draw[->] (3.75,4.4)--(3.75,3.1);	
	\node at (3.5, 3.8) {$b$};

	\draw[->] (0.75,2.9)--(0.75,1.6);	
	\node at (0.5, 2.3) {$t$};

	\draw[->] (2.75,2.9)--(2.75,1.6);	
	\node at (2.5, 2.3) {$t$};

	\draw[->] (4.75,3.9)--(4.75,1.6);	
	\node at (4.5, 2.8) {$t$};

	\draw[->] (6.75,3.9)--(6.75,1.6);	
	\node at (6.5, 2.8) {$t$};

	\draw[->] (0.75,1.4)--(0.75,0.1);	
	\node at (0.5,0.8) {$W$};

	\draw[->] (1.75,2.4)--(1.75,0.1);	
	\node at (1.5,1.4) {$W$};

	\draw[->] (4.75,1.4)--(4.75,0.1);	
	\node at (4.5,0.8) {$W$};

	\draw[->] (5.75,3.4)--(5.75,0.1);	
	\node at (5.5,1.8) {$W$};

	\node at (0.75, -0.5) {(a)};
	\node at (1.75, -0.5) {(b)};	
	\node at (2.75, -0.5) {(c)};
	\node at (3.75, -0.5) {(d)};	
	\node at (4.75, -0.5) {(e)};	
	\node at (5.75, -0.5) {(f)};		
	\node at (6.75, -0.5) {(g)};		
	\node at (7.75, -0.5) {(h)};		
	
 \end{tikzpicture}
\end{center}
\caption{The possible mass hierarchies for $\tilde{g}\rightarrow b\tilde{b}$, $\tilde{b}\rightarrow t\chi_1^\pm$.  The arrows denote which, if any, high energy particles are produced in the cascade decay.}
\label{fig:masshierarchy2}
\end{figure}
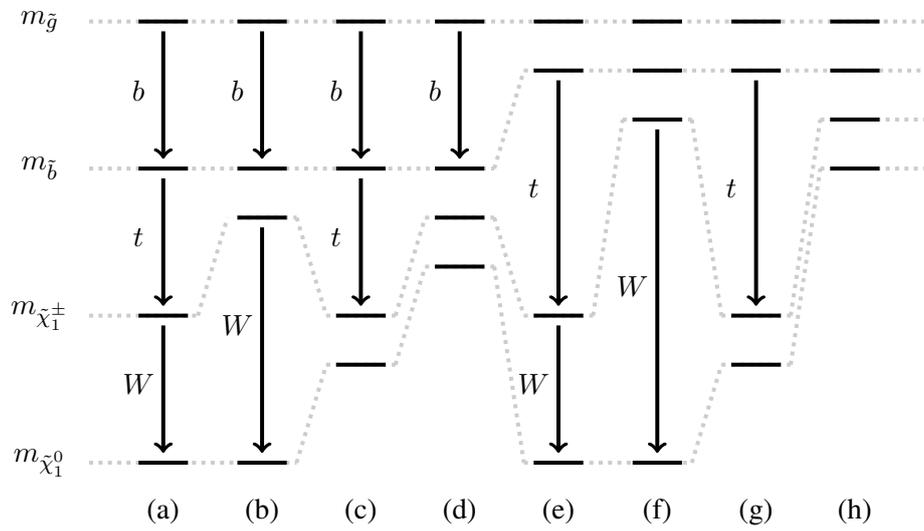

The flavor changing decays are $\tilde{t}\rightarrow b\chi^\pm_1$ and $\tilde{b}\rightarrow t\chi^\pm_1$.  Figures~\ref{fig:masshierarchy3} and~\ref{fig:masshierarchy2} schematically show the possible mass hierarchies for the stop and sbottom decays, respectively.  Many of the hierarchies are well-covered by traditional searches.  These include the scenarios in Figures~\ref{fig:masshierarchy3}a,b,c and~\ref{fig:masshierarchy2}a,c,e.  Other scenarios are already covered by the compressed models discussed in the body of the text, including Figures~\ref{fig:masshierarchy3}d and~\ref{fig:masshierarchy2}g.  The $b\bar{b}+E_T^\text{miss}$ equivalent (direct sbottom searches) equivalent of the $t\bar{t}+E_T^\text{miss}$ models (direct stop searches) discussed in the text include Figures~\ref{fig:masshierarchy3}g and~\ref{fig:masshierarchy2}d.  Two signatures which are not covered by direct stop/sbottom searches and are possibly uncovered by direct gluino searches include the $b\bar{b}WW$ and $WW$ (possibly without much $E_T^\text{miss}$) signatures in Figures~\ref{fig:masshierarchy3}e,f and~\ref{fig:masshierarchy2}b.  Dedicated study of such models, and the sensitivity of existing searches, is beyond the scope of this paper but should be part of the LHC search program in Run II and beyond.

\clearpage
\newpage

\bibliographystyle{unsrt}
\bibliography{paper}

\begin{thebibliography}{10}

\bibitem{Aad:2012tfa}
{\bf ATLAS}~Collaboration.
\newblock {Observation of a new particle in the search for the Standard Model
  Higgs boson with the ATLAS detector at the LHC}.
\newblock {\em Phys. Lett.}, B716:1--29, 2012.

\bibitem{Chatrchyan:2012ufa}
{\bf CMS}~Collaboration.
\newblock {Observation of a new boson at a mass of 125 GeV with the CMS
  experiment at the LHC}.
\newblock {\em Phys. Lett.}, B716:30--61, 2012.

\bibitem{Kramer:2012bx}
Michael Kramer et~al.
\newblock {Supersymmetry production cross sections in $pp$ collisions at
  $\sqrt{s}=7$ TeV}.
\newblock 2012.

\bibitem{Aad:2013wta}
{\bf ATLAS}~Collaboration.
\newblock {Search for new phenomena in final states with large jet
  multiplicities and missing transverse momentum at $\sqrt{s}=8$ TeV
  proton-proton collisions using the ATLAS experiment}.
\newblock {\em JHEP}, 1310:130, 2013.

\bibitem{Aad:2014lra}
{\bf ATLAS}~Collaboration.
\newblock {Search for strong production of supersymmetric particles in final
  states with missing transverse momentum and at least three $b$-jets at
  $\sqrt{s}$= 8 TeV proton-proton collisions with the ATLAS detector}.
\newblock {\em JHEP}, 1410:24, 2014.

\bibitem{Aad:2014pda}
{\bf ATLAS}~Collaboration.
\newblock {Search for supersymmetry at $\sqrt{s}=8$ TeV in final states with
  jets and two same-sign leptons or three leptons with the ATLAS detector}.
\newblock {\em JHEP}, 1406:035, 2014.

\bibitem{Aad:2014wea}
Georges Aad et~al.
\newblock {Search for squarks and gluinos with the ATLAS detector in final
  states with jets and missing transverse momentum using $\sqrt{s}=8$ TeV
  proton--proton collision data}.
\newblock {\em JHEP}, 1409:176, 2014.

\bibitem{Aad:2015mia}
Georges Aad et~al.
\newblock {Search for squarks and gluinos in events with isolated leptons, jets
  and missing transverse momentum at $\sqrt{s}=8$ TeV with the ATLAS detector}.
\newblock {\em JHEP}, 1504:116, 2015.

\bibitem{Chatrchyan:2014lfa}
{\bf CMS}~Collaboration.
\newblock {Search for new physics in the multijet and missing transverse
  momentum final state in proton-proton collisions at $\sqrt{s}$= 8 TeV}.
\newblock {\em JHEP}, 1406:055, 2014.

\bibitem{Chatrchyan:2013iqa}
{\bf CMS}~Collaboration.
\newblock {Search for supersymmetry in pp collisions at $\sqrt{s}=8$ TeV in
  events with a single lepton, large jet multiplicity, and multiple b jets}.
\newblock {\em Phys. Lett.}, B733:328--353, 2014.

\bibitem{Chatrchyan:2013fea}
{\bf CMS}~Collaboration.
\newblock {Search for new physics in events with same-sign dileptons and jets
  in pp collisions at $\sqrt{s}$ = 8 TeV}.
\newblock {\em JHEP}, 1401:163, 2014.

\bibitem{CMS-PAS-SUS-14-011}
{\bf CMS}~Collaboration.
\newblock {Exclusion limits on gluino and top-squark pair production in natural
  SUSY scenarios with inclusive razor and exclusive single-lepton searches at 8
  TeV.}
\newblock Technical Report CMS-PAS-SUS-14-011, CERN, Geneva, 2014.

\bibitem{CMS-PAS-SUS-13-016}
{\bf CMS}~Collaboration.
\newblock {Search for supersymmetry in pp collisions at sqrt(s) = 8 Tev in
  events with two opposite sign leptons, large number of jets, b-tagged jets,
  and large missing transverse energy.}
\newblock Technical Report CMS-PAS-SUS-13-016, CERN, Geneva, 2013.

\bibitem{CMS-PAS-SUS-13-008}
{\bf CMS}~Collaboration.
\newblock {Search for supersymmetry in pp collisions at sqrt(s) = 8 TeV in
  events with three leptons and at least one b-tagged jet}.
\newblock Technical Report CMS-PAS-SUS-13-008, CERN, Geneva, 2013.

\bibitem{Aad:2012ywa}
{\bf ATLAS}~Collaboration.
\newblock {Search for a supersymmetric partner to the top quark in final states
  with jets and missing transverse momentum at $\sqrt{s}=7$ TeV with the ATLAS
  detector}.
\newblock {\em Phys. Rev. Lett.}, 109:211802, 2012.

\bibitem{Aad:2012xqa}
{\bf ATLAS}~Collaboration.
\newblock {Search for direct top squark pair production in final states with
  one isolated lepton, jets, and missing transverse momentum in $\sqrt{s}=7$
  TeV $pp$ collisions using 4.7 $fb^{-1}$ of ATLAS data}.
\newblock {\em Phys. Rev. Lett.}, 109:211803, 2012.

\bibitem{Aad:2012uu}
{\bf ATLAS}~Collaboration.
\newblock {Search for a heavy top-quark partner in final states with two
  leptons with the ATLAS detector at the LHC}.
\newblock {\em JHEP}, 1211:094, 2012.

\bibitem{Aad:2014kra}
{\bf ATLAS}~Collaboration.
\newblock {Search for top squark pair production in final states with one
  isolated lepton, jets, and missing transverse momentum in $\sqrt s =$8 TeV
  $pp$ collisions with the ATLAS detector}.
\newblock {\em JHEP}, 1411:118, 2014.

\bibitem{Aad:2014bva}
{\bf ATLAS}~Collaboration.
\newblock {Search for direct pair production of the top squark in all-hadronic
  final states in proton-proton collisions at $\sqrt{s}=8$ TeV with the ATLAS
  detector}.
\newblock {\em JHEP}, 1409:015, 2014.

\bibitem{Aad:2014qaa}
{\bf ATLAS}~Collaboration.
\newblock {Search for direct top-squark pair production in final states with
  two leptons in pp collisions at $\sqrt{s}=8$ TeV with the ATLAS detector}.
\newblock {\em JHEP}, 1406:124, 2014.

\bibitem{Aad:2014mfk}
{\bf ATLAS}~Collaboration.
\newblock {Measurement of Spin Correlation in Top-Antitop Quark Events and
  Search for Top Squark Pair Production in pp Collisions at $\sqrt{s}=8$ TeV
  Using the ATLAS Detector}.
\newblock {\em Phys. Rev. Lett.}, 114(14):142001, 2015.

\bibitem{Aad:2014nra}
{\bf ATLAS}~Collaboration.
\newblock {Search for pair-produced third-generation squarks decaying via charm
  quarks or in compressed supersymmetric scenarios in $pp$ collisions at
  $\sqrt{s}=8$ TeV with the ATLAS detector}.
\newblock {\em Phys. Rev.}, D90(5):052008, 2014.

\bibitem{Chatrchyan:2013xna}
{\bf CMS}~Collaboration.
\newblock {Search for top-squark pair production in the single-lepton final
  state in pp collisions at $\sqrt{s}$ = 8 TeV}.
\newblock {\em Eur. Phys. J.}, C73(12):2677, 2013.

\bibitem{Khachatryan:2015wza}
{\bf CMS}~Collaboration.
\newblock {Searches for third generation squark production in fully hadronic
  final states in proton-proton collisions at sqrt(s)=8 TeV}.
\newblock 2015.

\bibitem{CMS-PAS-SUS-13-009}
{\bf CMS}~Collaboration.
\newblock {Search for top squarks decaying to a charm quark and a neutralino in
  events with a jet and missing transverse momentum}.
\newblock Technical Report CMS-PAS-SUS-13-009, CERN, Geneva, 2014.

\bibitem{Khachatryan:2014doa}
{\bf CMS}~Collaboration.
\newblock {Search for top-squark pairs decaying into Higgs or Z bosons in pp
  collisions at $\sqrt{s}=8$ TeV}.
\newblock {\em Phys. Lett.}, B736:371--397, 2014.

\bibitem{Chatrchyan:2014goa}
{\bf CMS}~Collaboration.
\newblock {Search for supersymmetry with razor variables in $pp$ collisions at
  $\sqrt{s}=7$ TeV}.
\newblock {\em Phys. Rev.}, D90(11):112001, 2014.

\bibitem{Chatrchyan:2012uea}
{\bf CMS}~Collaboration.
\newblock {Inclusive search for supersymmetry using the razor variables in $pp$
  collisions at $\sqrt{s}=7$ TeV}.
\newblock {\em Phys. Rev. Lett.}, 111(8):081802, 2013.

\bibitem{Agashe:2014kda}
K.A. Olive et~al.
\newblock {Review of Particle Physics}.
\newblock {\em Chin. Phys.}, C38:090001, 2014.

\bibitem{Bai:2012gs}
Yang Bai, Hsin-Chia Cheng, Jason Gallicchio, and Jiayin Gu.
\newblock {Stop the Top Background of the Stop Search}.
\newblock {\em JHEP}, 1207:110, 2012.

\bibitem{Graesser:2012qy}
Michael~L. Graesser and Jessie Shelton.
\newblock {Hunting Mixed Top Squark Decays}.
\newblock {\em Phys. Rev. Lett.}, 111(12):121802, 2013.

\bibitem{Nachman:2013bia}
Benjamin Nachman and Christopher~G. Lester.
\newblock {Significance Variables}.
\newblock {\em Phys. Rev.}, D88(7):075013, 2013.

\bibitem{aux2}
{\bf ATLAS}~Collaboration.
\newblock Susy-2013-15 auxillary material, 2014.

\bibitem{Alwall:2014hca}
J.~Alwall et~al.
\newblock {The automated computation of tree-level and next-to-leading order
  differential cross sections, and their matching to parton shower
  simulations}.
\newblock {\em JHEP}, 1407:079, 2014.

\bibitem{aux}
{\bf ATLAS}~Collaboration.
\newblock Susy-2013-09 auxillary material, 2014.

\bibitem{Read:2002hq}
Alexander~L. Read.
\newblock {Presentation of search results: The CL(s) technique}.
\newblock {\em J. Phys.}, G28:2693--2704, 2002.

\bibitem{Borschensky:2014cia}
Christoph Borschensky et~al.
\newblock {Squark and gluino production cross sections in pp collisions at
  $\sqrt s=$ 13, 14, 33 and 100 TeV}.
\newblock {\em Eur. Phys. J.}, C74(12):3174, 2014.

\bibitem{deSimone:2014pda}
Andrea De~Simone, Gian~Francesco Giudice, and Alessandro Strumia.
\newblock {Benchmarks for Dark Matter Searches at the LHC}.
\newblock {\em JHEP}, 1406:081, 2014.

\bibitem{Delgado:2012eu}
Antonio Delgado et~al.
\newblock {The light stop window}.
\newblock {\em Eur. Phys. J.}, C73(3):2370, 2013.

\bibitem{Profumo:2004at}
Stefano Profumo and Carlos~E. Yaguna.
\newblock {A Statistical analysis of supersymmetric dark matter in the MSSM
  after WMAP}.
\newblock {\em Phys. Rev.}, D70:095004, 2004.

\end{thebibliography}

\end{document}